\documentstyle[editedvolume,numreferences]{crckapb} 

\begin{opening} 
\title{ Properties of nuclei in the neutron star crust} 
 
\author{P. Magierski$^1$, A. Bulgac$^2$, P.-H. Heenen$^3$} 
\institute{$^1$Physics Faculty, Warsaw University of Technology\\ 
           ul. Koszykowa 75, PL--00662, Warsaw, Poland \\ 
           $^2$Department of Physics, University of Washington\\ 
           P.O. Box 351560, Seattle, WA 98195--1560, USA\\ 
           $^3$Service de Physique Nucl\'{e}aire Th\'{e}orique\\ 
           U.L.B - C.P. 229, B 1050 Brussels, Belgium} 

\end{opening}

\begin{document} 
 

\begin{abstract} 
In the present study we investigate the static properties of
nuclei in the inner crust of neutron stars.
Using the Hartree-Fock method in coordinate
space, together with the
semiclassical approximation, we examine the patterns of phase
transitions. 
\end{abstract}

\bigskip

The properties of the outer parts of neutron stars are important for the
understanding of several observational issues.
Despite the fact that the crust is typically less than 10\% of the
radius of the star one expects that such phenomena as: thermal X-ray emission
from the stellar surface, X-ray burst sources, or
the sudden speed-ups in the rotation rate of
some neutron stars may be related to the structure of the crust.

In the inner crust of the star, the increasing pressure and density
force nuclei to loose some of their neutrons. Hence deeper in the star,
the nuclei are immersed in a neutron gas. Since nuclei in such an
environment cannot be explored experimentally, our understanding of matter
under these conditions is based on modelling. An agreement
has been reached in the literature concerning the existence of the following
chain of phase transitions as the density increases: nuclei
$\rightarrow$ rods $\rightarrow$ plates $\rightarrow$ tubes $\rightarrow$
bubbles $\rightarrow$ uniform nuclear matter \cite{pra}.
The appearance of different phases is attributed to the interplay
between Coulomb and surface energies. Most published
works are based on the minimization of some density functional in a
single Wigner--Seitz cell and neglect the shell effects. The
Hartree--Fock calculations performed so far are limited to
Wigner--Seitz cells with geometries reflecting the
nuclear shape and not the lattice structure \cite{negele,dha}.
Hence the shell effects associated with the scattering of
unbound neutrons on nuclear inhomogeneities were not
properly taken into account.

There is no well
established terminology for the energy corrections we are considering
here, even though the problem has been addressed to some extent
by other authors. In the case of finite systems, the energy difference
between the true binding energy and the liquid drop energy of a given
system is typically referred to as shell correction energy. In field
theory a somewhat similar energy appears, due to various fluctuation
induced effects and it is generically referred to as the Casimir
energy \cite{casimir}:
\begin{equation}
E_{Casimir} = \frac{1}{2}\int _{-\infty }^\infty d\varepsilon
\varepsilon [g(\varepsilon ,{\bf l} )-g_0(\varepsilon )],
\end{equation}
where $g_0 (\varepsilon )$ is the density of states per unit volume
for the fields in the absence of any objects, $g (\varepsilon ,{\bf
l})$ is the density of states per unit volume in the presence of some
``foreign'' objects, such as plates, spheres, etc., and ${\bf l}$ is an
ensemble of geometrical parameters describing these objects and their
relative geometrical arrangement. A similar formula can be written for
neutron matter energy \cite{bma1,bwi}
\begin{equation}
E_{nm}=
\int_{-\infty}^ \mu    d\varepsilon\varepsilon g  (\varepsilon ,{\bf l})
-\int_{-\infty}^{\mu _0}d\varepsilon\varepsilon g_0(\varepsilon ,{\bf l}),
\end{equation}
with the notable difference in the upper integration limit.  In the
above equation $g _0(\varepsilon , {\bf l})$ stands for the
Thomas--Fermi or liquid drop density of states of the inhomogeneous
phase and $g (\varepsilon ,{\bf l} )$ is the true quantum density of
states in the presence of inhomogeneities.  The parameters: $\mu$ and
$\mu _0$ are determined from the requirement that the system has a
given average density:
\begin{equation}
\rho =\int _{-\infty }^\mu d\varepsilon  g (\varepsilon ,{\bf l} )
=  \int _{-\infty }^{\mu _0} d\varepsilon g _0(\varepsilon ,{\bf l} ).
\end{equation}
Since in infinite matter the presence of various inhomogeneities does
not lead to the formation of discrete levels, the effects we
shall consider here arise from the scattering
states, which is in complete analogy with the procedure for computing
the Casimir energy.

We shall only consider here spherical bubble--like, rod--like and plate--like
phases. One can distinguish two types of
``bubbles'': {\it i)} nuclei--like structures embedded in a neutron
gas and {\it ii)} void--like structures. By voids we mean the regions
in which the nuclear density is significantly lower than in the
surrounding space.  In the first case {\it i)}, the single particle
wave functions can be roughly separated into  two classes, those
localized mostly inside the nuclei--like structures and those which
are completely delocalized. A fermion in a delocalized state will
spend some time inside the ``nuclei'' too, but since the potential
experienced by a nucleon is deeper there, the local momentum is larger
and thus the relative time and relative probability to find a nucleon
in this region is smaller \cite{bcf}. One can approximately replace then the
``nuclei'' by an effective repulsive potential of roughly the same
shape. In the case of a ``bubble'', when the probability to find a
nucleon inside a ``bubble'' is reduced, again such an approximation
appears reasonable. The same holds true for rod--like or
plate--like structures. There are of course a number of ``resonant''
delocalized states, whose amplitudes behave in an opposite manner. 
However, the number of such
``resonant'' states is small and brings only small corrections.
In all these phases the shell effects depend on the
structure and stability of periodic orbits in the system
\cite{brack}. 

In the semiclassical approximation
it can be shown that the leading contribution
to the shell energy (which may be thought of as the
interaction energy),
for two obstacles being either spherical ($i=0$),
rod--like ($i=1$) or slab--like ($i=2$)
nuclei located at a distance $d$ is given by \cite{phe}:
\begin{equation} \label{shell}
E_{shell} \approx
\frac{\hbar^{2} L^{i} R^{2-i}}{8m_{n}} \left ( \frac{3}{\pi}
\right )^{\frac{2+i}{6}}
\frac{(\rho_n^{out})^{\frac{2+i}{6}}}{d^{\frac{6-i}{2}}}
\cos\left (2k^{n}_{F}d-i\frac{\pi}{4}\right ) ,
\end{equation}
where $m_{n}$ is the neutron
mass, $k_{F}^{n}$ and $\rho_{n}^{out}$ denote the Fermi momentum
and density of unbound neutrons, respectively\footnote{
In the Refs.\cite{bma1,bma2} the factor ``$1/\pi$'' has been
missed in the expression for the
shell energy of rod--like phase.
However it does not change the 
conclusions.}. 
We have assumed that rods and slabs are parallel to each other.
In this equation, $L$ defines the length of the obstacle, and $R$ 
its radius (in the case of a slab it is defined
as half of its width). It turns out
that the semiclassical estimations yield energy corrections
of the same order of magnitude as the
liquid drop energy differences between the various phases of the inhomogeneous
nuclear matter \cite{bma1,bma2}. It has lead to the hypothesis that
the inner crust may have a quite complicated structure, probably even
completely disordered \cite{bma1}.

Unfortunately the semiclassical approach
has some drawbacks. First, it assumes that the obstacles are inpenetrable
scatterers which may overestimate the amplitude of the shell energy.
Second, the method lacks of the mutual interplay between the shell energy
and the liquid drop energy. Such a coupling is quite important since it
allows a part of the shell energy to be ``absorbed'' into deformation.
Hence a microscopic treatment of the problem is needed where both effects,
i.e. the one coming from the liquid drop energy (where mainly the surface and
Coulomb terms are involved) and the shell energy term are treated on the
same footing. Therefore we have applied the Hartree-Fock 
(HF) approach in the coordinate
space with a Skyrme effective interaction (see \cite{phe,bfh} for
details).
The analysis of coexistence and stability of different nuclear phases
requires construction of an adiabatic path between different configurations.
It has lead us to abandon the Wigner-Seitz approximation 
for the Coulomb interaction, solving
instead the Poisson equation for the electric potential
within the box with periodic boundary
conditions, the total charge of the box being zero.
In our approach the electrons are assumed to
form a uniform relativistic gas where screening effects are neglected.

Summarizing, we minimize an energy functional
for the neutron-proton-electron matter within the box. The energy
functional describes both the
liquid drop part of the energy and the shell effects. Namely:
\begin{equation} \label{energy}
E_{tot}(\rho_{p},\rho_{n},\rho_{e})=
E_{TF}(\rho_{p},\rho_{n},\rho_{e}) +
E^{in}_{shell}(\rho^{in}_{p},\rho^{in}_{n}) +
E^{out}_{shell}(\rho^{out}_{n},\rho^{out}_{p}),
\end{equation}
where $\rho_{n}, \rho_{p}, \rho_{e}$ denote the density of neutrons,
protons and electrons, respectively. The charge neutrality condition
implies $\rho_{e}=\rho_{p}$. The total nuclear density
$\rho=\rho_{n}+\rho_{p}$. In the above expression $E_{TF}$ represents
the Thomas-Fermi energy of the system
obtained by constructing the energy functional for a Skyrme interaction.
We have used the SLy4 force \cite{chab} which
give a correct description of neutron matter at low densities
\cite{dha}.

The shell energy consists of two parts. The one denoted by
$E_{shell}^{in}$ is determined by the density of neutrons $\rho_{n}^{in}$
and protons $\rho_{p}^{in}$ inside the nucleus, whereas
$E_{shell}^{out}$  is a
function of unbound neutrons $\rho_{n}^{out}$ and
protons $\rho_{p}^{out}$.
Since $\rho_{p}^{out}=0$ for
$\rho < 0.08$ fm$^{-3}$ \cite{dha}, one can neglegt the
contribution to the energy due to the unbound protons.

The HF calculations confirm the importance of the shell effects
associated with unbound neutrons \cite{phe}. In particular, they 
show that the shell
effects associated with  unbound neutrons
may reverse the phase transition order predicted by the
liquid drop based approaches.  Moreover, the number of phase
transitions increases since the same phase may appear
for various density ranges. Our results suggest also that the
number of different phases present in the crust is larger since
one has to take into account various lattice geometries
which may compete. This important fact was disregarded in 
earlier investigations since in liquid drop based approaches
the system favours only one lattice type for a given nuclear shape.
However it is no longer the case  when one considers the neutron shell
energy which is very sensitive to the spatial order in the system.

In Fig.1, we  show several nuclear configurations that
can be formed at subnuclear densities. One notices the
presence of different types of crystal lattices for the same
nuclear shapes. The integrated densities are defined by:
\begin{eqnarray}
\rho_{i}^{z}(x,y)&=&\int_{0}^{d} \rho_{i}(x,y,z)dz, \\
\rho_{i}^{y}(x,z)&=&\int_{0}^{d} \rho_{i}(x,y,z)dy, 
\end{eqnarray}
where $i=n,p$ and $d$ denotes the size of the box.
\begin{figure} 
\vspace{20cm}  
\caption{The integrated densities (see text
for definition) for various
ground state or isomeric nuclear configurations
at constant density and $Z/A$ ratio. The size
of the box was equal to $d=26 fm$ - subfigures a) and b),
$d=23.4 fm $ - subfigures c) and d), $d=20.8 fm$
-subfigures e) and f). 
The spherical nuclear phase is
shown in the subfigures a) (scc) and 
d) (bcc), rod--like phase -- subfigures  b) and c),
slab--like phase -- subfigure e), bubble--like
phase: subfigure f). Note different lattice types
for the same nuclear shape.} 
\end{figure} 

Once a phase is formed, there is a positional order maintained by the
Coulomb repulsion between spherical nuclei, rods or slabs.
The Coulomb energy is a
smooth function of the void displacement but not the shell energy.
Since several orbits contribute to the shell
effects (except for the slab--like phase), the displacement of a single
bubble--like or rod--like void from its equilibrium position in the
lattice will give rise to interference effects \cite{bma1}.

The main conclusion of the present work is that the amplitude
of the shell effects is comparable to the energy differences
between the various phases determined in simpler liquid drop type models.
Our results suggest that the inhomogeneous phase has a
complicated structure, maybe  completely disordered,
with several types of coexisting shapes.
The few results highlighted here shows
the richness of these systems where both static and dynamic properties
are challenging to describe.

\begin{center}
{\bf ACKNOWLEDGMENTS}
\end{center}

This research was supported in part by the Polish Committee
for Scientific Research (KBN) and the Wallonie/Brussels-Poland
integrated action program. Authors would like to thank
Hubert Flocard for providing us with the numerical code
solving the Poisson equation. We also appreciate many discussions
with A. Wirzba, Y. Yu, S.A. Chin and H. Forbert.

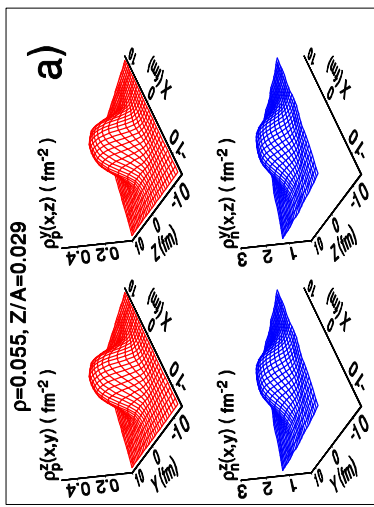
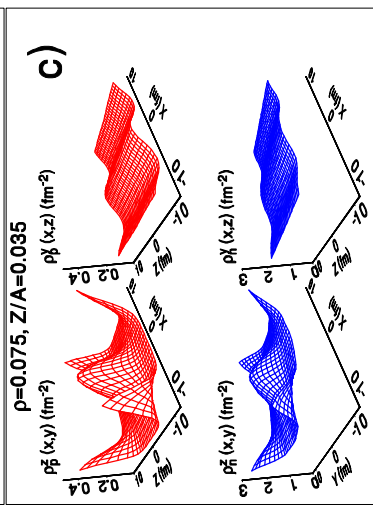
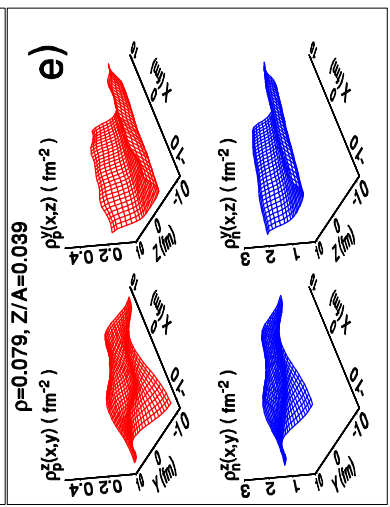
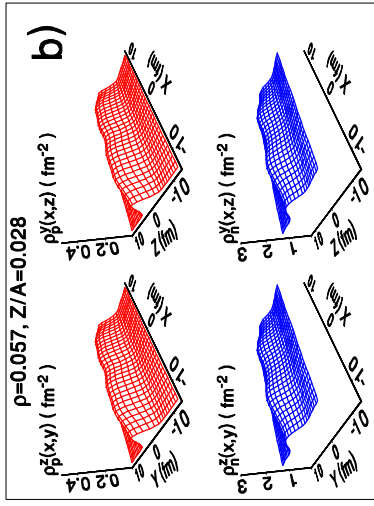
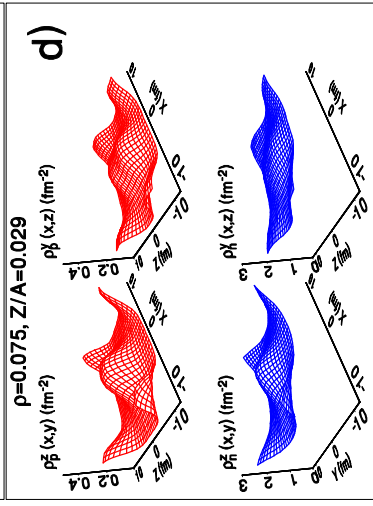
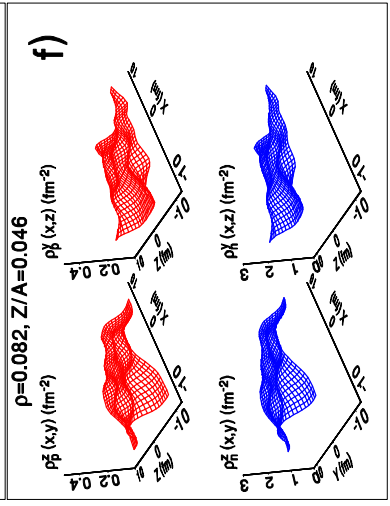
\end{document}